# ON IDENTIFIABLE POLYTOPE CHARACTERIZATION FOR POLYTOPIC MATRIX FACTORIZATION

*Bariscan Bozkurt and Alper T. Erdogan*

EE Department and KUIS AI Lab, Koc University, Istanbul

**ABSTRACT**

Polytopic matrix factorization (PMF) is a recently introduced matrix decomposition method in which the data vectors are modeled as linear transformations of samples from a polytope. The successful recovery of the original factors in the generative PMF model is conditioned on the "identifiability" of the chosen polytope. In this article, we investigate the problem of determining the identifiability of a polytope. The identifiability condition requires the polytope to be permutation- and/or-sign-only invariant. We show how this problem can be efficiently solved by using a graph automorphism algorithm. In particular, we show that checking only the generating set of the linear automorphism group of a polytope, which corresponds to the automorphism group of an edge-colored complete graph, is sufficient. This property prevents checking all the elements of the permutation group, which requires factorial algorithm complexity. We demonstrate the feasibility of the proposed approach through some numerical experiments.

## 1. INTRODUCTION

Polytopic Matrix Factorization (PMF) was recently introduced as a structured matrix decomposition approach [1]. The PMF framework models the columns of the input matrix as linear transformations of some latent vectors drawn from a polytope. The choice of the polytope reflects assumptions on the attributes of the latent components and their relationships. There are infinitely many so-called "identifiable polytopes" to choose from, enabling latent vectors with heterogeneous structures, such as a mixture of signed and nonnegative components and sparse subvectors. However, not all convex polytopes guarantee identifiability. In order for a polytope to be identifiable, its vertex set should be invariant only to linear transformations corresponding to permutations and/or sign alterations of the components [2]. In this article, we address efficient approaches for polytope identifiability based on the algebraic structure of the polytope symmetries.

This work is partially supported by an AI Fellowship provided by the KUIS AI Lab.

Combinatorial and geometric symmetries of polytopes have been of great interest due to their mathematical implications as well as their applications in engineering and optimization problems. Computing the linear, projective, and combinatorial symmetries of polyhedra are investigated in [3, 4]. Moreover, the symmetry groups of polyhedra play an essential role in linear, and integer linear programming [3, 5, 6]. In [7], the symmetry groups of linear systems with bounded polyhedral constraints are studied to find the state-space and input-space automorphisms. In this paper, we consider the linear automorphism groups of polytopes to characterize the identifiability for Polytopic Matrix Factorization. In particular, we show that it is sufficient to search only the generating set of the linear automorphism group whose cardinality is less than the number of vertices of the polytope instead of checking all possible permutations of the vertices, which has factorial complexity. Furthermore, we can obtain the generative set members using existing efficient graph automorphism algorithms.

The following is the outline of the article: Section 2 provides the identifiability definition and its algebraic consequences. In Section 3, we give an algorithmic framework to characterize the identifiability using a graph automorphism algorithm. Section 4 provides our numerical experiments and conclusion.

## 2. POLYTOPIC MATRIX FACTORIZATION AND IDENTIFIABILITY

The representation of a nonempty polyheral set in $\mathbb{R}^n$ of the form $\{x \in \mathbb{R}^n | Ax \preceq b\}$ is called an H-representation. A bounded polyhedron is called a polytope. By Minkowski-Weyl theorem [8], there exist two equivalent H and V representations of a given polytope $P$, i.e.,

$$P = \{x \in \mathbb{R}^n | Ax \preceq b\} \iff P = conv\{v_1, \ldots, v_m\}$$

where the set $\{v_1, \ldots, v_m\}$ is the set of all vertices of $P$, and $conv\{.\}$ is the convex hull of a given set. We will refer to the matrix $V = \begin{bmatrix} v_1 & v_2 & \ldots & v_m \end{bmatrix} \in \mathbb{R}^{n \times m}$ as the vertex matrix of the polytope.

Throughout the article, we will assume that the vertex matrix $V$ of a given polytope $P$ is full column rank, i.e.,

$rank(\boldsymbol{V}) = n$, and $n \leq m$. Therefore, the set $\{\boldsymbol{v}_1, \ldots, \boldsymbol{v}_m\}$ spans $\mathbb{R}^n$, and can be reduced to a basis.

In PMF, input data matrix $\boldsymbol{Y}$ is assumed to be of the form $\boldsymbol{Y} = \boldsymbol{HS}$ where $\boldsymbol{H}$ is a full column-rank matrix and the columns of $\boldsymbol{S}$ are in the polytope $P$. The identifiability of the original factors from $\boldsymbol{Y}$, up to some scaling and permutation ambiguities is condition on the identifiability of $P$ [2], which is described by the following definition:

**Definition 2.1.** Let $P \subset \mathbb{R}^n$ be a given polytope with $P = conv\{\boldsymbol{v}_1, \ldots, \boldsymbol{v}_m\}$. Also, let $\boldsymbol{V} \in \mathbb{R}^{n \times m}$ represent its vertex matrix. A polytope is identifiable [2] if and only if for any permutation matrix $\boldsymbol{\Pi} \in \mathbb{R}^{m \times m}$, the solution to the equation

$$\boldsymbol{GV} = \boldsymbol{V}\boldsymbol{\Pi} \tag{1}$$

holds only for $\boldsymbol{G} = \boldsymbol{D}\bar{\boldsymbol{\Pi}}$, where $\boldsymbol{D} \in \mathbb{R}^{n \times n}$ is a diagonal matrix with entries $\{+1, -1\}$, and $\bar{\boldsymbol{\Pi}} \in \mathbb{R}^{n \times n}$ is a permutation matrix.

The matrices of the form $\boldsymbol{G} = \boldsymbol{D}\bar{\boldsymbol{\Pi}}$ are referred as the signed permutation matrices when $\boldsymbol{D}$ is a diagonal matrix with entries $\pm 1$. Observe that, any solution matrix $\boldsymbol{G}$ to (1) only permutes the vertices, and it is not hard to prove that the linear transformation of the convex polytope $P$ with $\boldsymbol{G}$ is itself, i.e., $\boldsymbol{G}(P) = P$.

Now, we present fundamental observations about the matrices satisfying (1).

**Lemma 2.1.** *For a given $\boldsymbol{\Pi} \in \mathbb{R}^{m \times m}$, if $\boldsymbol{G}$ satisfying (1) exists then it is unique.*

*Proof.* Let $\boldsymbol{G}$ and $\boldsymbol{G}'$ be two solutions of (1). Then we can write,

$$\begin{aligned} \boldsymbol{GV} &= \boldsymbol{V}\boldsymbol{\Pi} = \boldsymbol{G}'\boldsymbol{V} \\ \boldsymbol{GV} - \boldsymbol{G}'\boldsymbol{V} &= \boldsymbol{0}_{n \times m} \\ (\boldsymbol{G} - \boldsymbol{G}')\boldsymbol{V} &= \boldsymbol{0}_{n \times m}. \end{aligned}$$

The full column rank property for $\boldsymbol{V}$ implies $\boldsymbol{G} - \boldsymbol{G}' = \boldsymbol{0}_{n \times n}$. Therefore, for a fixed $\boldsymbol{\Pi}$, $\boldsymbol{G}$ matrix satisfying (1) is unique. ∎

**Lemma 2.2.** *For any solution $\boldsymbol{G}$ of (1), $det(\boldsymbol{G}) = \pm 1$.*

*Proof.* Let $\boldsymbol{G} \in \mathbb{R}^{n \times n}$, $\boldsymbol{\Pi} \in \mathbb{R}^{m \times m}$ such that $\boldsymbol{GV} = \boldsymbol{V}\boldsymbol{\Pi}$. Multiplying both sides by $\boldsymbol{G}$ from left, we observe that

$$\boldsymbol{G}^2\boldsymbol{V} = \boldsymbol{GV}\boldsymbol{\Pi} = \boldsymbol{V}\boldsymbol{\Pi}^2.$$

For any permutation matrix $\boldsymbol{\Pi} \in \mathbb{R}^{m \times m}$, there exist an integer $k \leq m$ [9], such that $\boldsymbol{\Pi}^k = \boldsymbol{I}$, where $\boldsymbol{I} \in \mathbb{R}^{m \times m}$ is the identity matrix. Then,

$$\boldsymbol{G}^k\boldsymbol{V} = \boldsymbol{V}\boldsymbol{\Pi}^k = \boldsymbol{V}.$$

Since $\boldsymbol{V}$ is full column rank, $\boldsymbol{G}^k = \boldsymbol{I}$.

$$det(\boldsymbol{G}^k) = 1 \iff det(\boldsymbol{G})^k = 1 \iff det(\boldsymbol{G}) = \pm 1.$$
∎

Lemma 2.2 also follows from the fact that a linear transformation that doesn't change the shape of the polytope also preserves its volume.

Before we provide some results about the group structure of the set of all matrices satisfying (1), we provide the definition of Group for the clarity of presentation [9]:

**Definition 2.2.** A group is a set $\mathscr{G}$ equipped with a binary operation $*$ from $\mathscr{G} \times \mathscr{G} \longrightarrow \mathscr{G}$ satisfying the following four axioms:

(A1) $\forall x, y, z \in \mathscr{G}$ $(x * y) * z = x * (y * z)$ (associative),

(A2) $\exists e \in \mathscr{G}$ s.t. $\forall x \in \mathscr{G}$, $e * x = x * e = x$, (identity exist),

(A3) $\forall x \in \mathscr{G}$, $\exists x' \in \mathscr{G}$ s.t. $x * x' = e = x' * x$, (inverse exists),

(A4) $\forall x, y \in \mathscr{G}$, $x * y \in \mathscr{G}$ (closure).

**Lemma 2.3.** *The set of all $\boldsymbol{G}$ matrices satisfying (1) together with matrix multiplication as the binary operation forms a group.*

*Proof.* Let us denote the set of all such matrices by $\mathscr{G}(P)$ for a given polytope $P$. Below we show that $\mathscr{G}(P)$ satisfies all the group axioms in Definition 2.2

(A1) The set satisfies associativity axiom since matrix multiplication is associative.

(A2) The identity matrix is the identity element of this set.

(A4) Let $\boldsymbol{G}_1, \boldsymbol{G}_2 \in \mathscr{G}(P)$, such that

$$\boldsymbol{G}_1\boldsymbol{V} = \boldsymbol{V}\boldsymbol{\Pi}_1, \text{ and } \boldsymbol{G}_2\boldsymbol{V} = \boldsymbol{V}\boldsymbol{\Pi}_2.$$

Then we can write,

$$\boldsymbol{G}_1\boldsymbol{G}_2\boldsymbol{V} = \boldsymbol{G}_1\boldsymbol{V}\boldsymbol{\Pi}_2 = \boldsymbol{V}\boldsymbol{\Pi}_1\boldsymbol{\Pi}_2.$$

Since $\boldsymbol{\Pi}_1\boldsymbol{\Pi}_2$ is another permutation matrix, $\boldsymbol{G}_1\boldsymbol{G}_2 \in \mathscr{G}(P)$. $\boldsymbol{G}_2\boldsymbol{G}_1 \in \mathscr{G}(P)$ similarly follows.

(A3) $\forall \boldsymbol{G} \in \mathscr{G}(P), \boldsymbol{G}^{-1} \in \mathscr{G}(P)$ follows from the proofs of Lemma 2.2 and (A4).

∎

We refer to the group $\mathscr{G}(P)$ as the linear automorphism group of polytope $P$ since each element in $\mathscr{G}(P)$ is a linear map that transforms $P$ to itself. Moreover, due to the relation in (1), $\mathscr{G}(P)$ is isomorphic to a subgroup of the permutation group $S_m$ of $m \times m$ permutation matrices. As a result, $\mathscr{G}(P)$ is a finite group that is isomorphic to a subgroup of $S_m$.

A generating set $Gen(\mathscr{G})$ of a group $\mathscr{G}$ is defined as the subset of the group such that each group element can be expressed as the product of the elements of $Gen(\mathscr{G})$ and their inverses. The following theorem shows that the problem of characterizing the identifiability of polytope $P$ can be reduced to determining whether all the elements of the generating set of the group $\mathscr{G}(P)$ are sign/permutation matrices.

**Theorem 2.4.** *Let $Gen(\mathscr{G}(P)) = \{G_1, G_2, \ldots, G_r\}$ be the generating set of linear automorphisms for a given polytope P. If all the elements of $Gen(\mathscr{G}(P))$ are signed permutation matrices, then P is identifiable.*

*Proof.* We present the sketch of the proof, but the idea simply follows by induction. Let $\{\Pi_1, \Pi_2, \ldots, \Pi_r\}$ represent the corresponding permutation matrices for $Gen(\mathscr{G}(P))$, i.e., $G_i V = V \Pi_i$ for all $i = 1, \ldots, r$. Now consider the pair $G_1$ and $G_2$, and assume that both are signed permutation matrices, i.e., $\exists\, D_1, \bar{\Pi}_1, D_2, \bar{\Pi}_2 \in \mathbb{R}^{n \times n}$ such that

$$G_1 = D_1 \bar{\Pi}_1, \quad G_2 = D_2 \bar{\Pi}_2.$$

Based on (1), we obtain

$$G_1 G_2 V = G_1 V \Pi_2 = V \Pi_1 \Pi_2.$$

Therefore, $\Pi_1 \Pi_2$ is the permutation matrix corresponding to $G_1 G_2$. Moreover, writing the product of the pair $G_1, G_2$ explicitly,

$$\begin{aligned} G_1 G_2 &= D_1 \bar{\Pi}_1 D_2 \bar{\Pi}_2 \\ &= D_1 \bar{\Pi}_1 D_2 \bar{\Pi}_1^T \bar{\Pi}_1 \bar{\Pi}_2 \\ &= D_1 D_2' \bar{\Pi}_1 \bar{\Pi}_2, \end{aligned}$$

where $D_2' = \bar{\Pi}_1 D_2 \bar{\Pi}_1^T$ is another diagonal matrix. Therefore, $G_1 G_2$ is a signed permutation matrix since $D_1 D_2'$ is a diagonal and $\bar{\Pi}_1 \bar{\Pi}_2$ is a permutation matrix. The rest of the proof follows from induction since each element of $\mathscr{G}(P)$ can be represented as the product of the generating set members. ∎

## 3. IDENTIFIABLE POLYTOPE CHARACTERIZATION

Determining the automorphism group of a convex polytope and its generating set is a hard problem [3, 4]. As a solution approach, for a given polytope $P$ with $m$ vertices, we can construct an edge-colored complete graph $G_P$ with $m$ nodes, where each node corresponds to a vertex in $P$ such that the automorphism group of $P$ is isomorphic to the automorphism group of this graph [3, 4]. Moreover, there are different coloring techniques of the edge-graph of a polytope such that the combinatorial automorphism groups of the corresponding colored edge-graphs are isomorphic to the linear or orthogonal symmetries of the polytope [10]. We will follow the edge-colored complete graph construction from a polytope to calculate the generating set of the linear automorphism group of the polytope as described in [3, 4]. For this purpose, let $P \in \mathbb{R}^n$ be the given polytope with the vertex matrix $V = [v_1\ v_2\ \ldots\ v_m] \in \mathbb{R}^{n \times m}$. Also, let $Q = VV^T \in \mathbb{R}^{n \times n}$ and $C = V^T Q^{-1} V \in \mathbb{R}^{m \times m}$. For the given polytope $P \in \mathbb{R}^n$ with $m$ vertices, we set up an edge-colored complete graph $G_P$ with m nodes where each

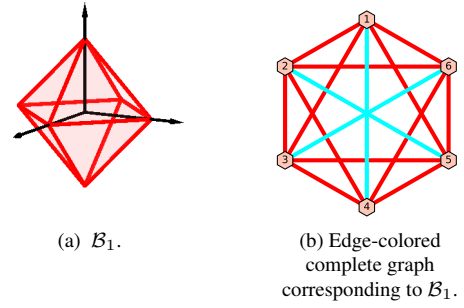

(a) $\mathcal{B}_1$.  (b) Edge-colored complete graph corresponding to $\mathcal{B}_1$.

**Fig. 1**: $\mathcal{B}_1$ and Corresponding $G_P$

pair of distinct nodes are connected (complete graph), and the edge color from $i^{th}$ node to $j^{th}$ node is given by the inner product $v_i^T Q^{-1} v_j = C_{i,j}$ for $i \neq j$ (edge-colored graph). The matrix $C$ is called the colored (or weighted) adjacency matrix of the graph $G_P$. For example, Figure 1 illustrates the $\ell_1$-norm ball ($\mathcal{B}_1$) in three dimensions and its corresponding edge-colored complete graph. We simply construct a complete graph with 6 nodes, as shown in Figure 1.(b), where each pair of distinct vertices are connected. For $\mathcal{B}_1 \in \mathbb{R}^3$ in Figure 1.(a), we can write the vertex matrix as $V = \begin{bmatrix} I & -I \end{bmatrix}$, and its corresponding coloring matrix $C$ as

$$C_{i,j} = \begin{cases} 0.5 & i = j, \\ -0.5 & |i - j| = 3, \\ 0 & \text{otherwise.} \end{cases} \quad (2)$$

Therefore, in Figure 1.(b), the edge from node-1 to node-4 is represented as blue and the other vertices connected to node-1 as red since $C_{1,4} = -0.5$ and $C_{1,j} = 0$ for $j = 2, 3, 5, 6$. The colors of the other vertices are represented similarly. For $G_P$, the permutation matrices which satisfy $C = \Pi^T C \Pi$ are the automorphisms of the graph, which are bijections from its nodes to itself preserving the colored adjacency.

As we will algebraically prove in Theorem 3.1, the linear automorphism group of a given polytope $P$ and the automorphism group of the constructed $G_P$ are isomorphic. The graph automorphism problem, which is the deciding the graph isomorphisms from graph to itself, is in the complexity class of NP, and it is not known whether it is NP-complete or not [11, 12]. However, there are well-known graph isomorphism algorithms that are practically used in many applications, although the worst-case run time of these algorithms is exponential [3, 4, 11, 12]. For the identifiable polytope characterization problem, we adapted the graph automorphism algorithm in [12]. The numerical examples in Section 4 demonstrate that it works much more efficiently than the brute-force search for all possible permutations of vertices, which has factorial run-time. Moreover, since the graph automorphism algorithm gives us the generating set of the automorphism group with at most $(m - 1)$ elements for $m$ vertices [12, 7], the complexity of the proposed algorithm for polytope identi-fiability characterization will be essentially equivalent to the complexity of graph isomorphism problem.

We now provide the algebraic proof that the linear automorphism group of a given polytope $P$ and the automorphism group of the edge-colored complete graph $G_P$ is isomorphic. For more graph-theoretical proofs and explanations, we refer to [3, 4]. Finally, combining this observation with Theorem 2.4, we will propose the identifiability characterization of a polytope.

**Theorem 3.1.** *Let $\Pi \in \mathbb{R}^{m \times m}$ be a permutation matrix. Then,*
$$GV = V\Pi \iff C = \Pi^T C \Pi$$
*In that case, $G = V\Pi V^T Q^{-1} = V\Pi V^\dagger$.*

*Proof.* ($\Rightarrow$ "if part") Assume $GV = V\Pi$. We note,
$$GQG^T = GVV^T G^T = V\Pi\Pi^T V^T = VV^T = Q.$$

Therefore, $G^T Q^{-1} G = Q^{-1}$. Then, we obtain
$$\Pi^T C \Pi = \Pi^T V^T Q^{-1} V\Pi = V^T G^T Q^{-1} GV$$
$$= V^T Q^{-1} V = C.$$

($\Leftarrow$ "only if part") Assume $C = \Pi^T C \Pi$ for a permutation matrix. Let $G = V\Pi V^T Q^{-1}$. Then,
$$GV = V\Pi V^T Q^{-1} V = V\Pi C = VC\Pi$$
$$= VV^T Q^{-1} V\Pi = V\Pi.$$
∎

Algorithm 1 summarizes the proposed algorithmic approach based on Theorem 2.4 and Theorem 3.1.

## 4. NUMERICAL EXAMPLES AND CONCLUSION

To implement the proposed Algorithm 1, we used *SageMath* [13] software in *Python3.7* for graph automorphism, which implements the algorithm in [12]. To experiment with practical polytopes for PMF setting, we generated random polytopes similar to the example given in [2]. For any dimension $n \geq 3$, we first randomly decide for each component to be non-negative, i.e., $s_j \in [0, 1]$, or signed, i.e., $s_j \in [-1, 1]$. Then, we select $q$ random subvectors to apply sparsity constraints in the form $\left\| \begin{bmatrix} s_{j_1^{(i)}} & s_{j_2^{(i)}} & \ldots & s_{j_{l_i}^{(i)}} \end{bmatrix} \right\|_1 \leq 1$, where $q \in \{2, \ldots, n\}$ is randomly selected, $i \in \{1, \ldots, q\}$, $l_i \in \{2, \ldots, n\}$ is the random length of the subvector-$i$ and $\{j_1^{(i)}, \ldots, j_{l_i}^{(i)}\}$ are the randomly selected indices of the subvector-$i$. The number of relative sparsity constraint is selected randomly. In this way, we obtain the H-representation of the polytope. Then, we calculate the vertices of the generated polytope. We generated a practical polytopes dataset in this way which contains 6000 randomly generated polytopes up to dimension 10.

**Algorithm 1** Identifiability Decision
    **Input**: $V = [v_1 \ v_2 \ \ldots \ v_m] \in \mathbb{R}^{n \times m}$ (vertex matrix)
    **Output**: *Identifiable* (Boolean)
1:  $Q = VV^T$
2:  $C = V^T Q^{-1} V$
3:  Construct an edge-colored complete graph $G_P$ with the edge colors from $C$.
4:  Calculate $Gen(Aut(G_P))$, the generator set of the automorphism group of $G_P$ with a graph isomorphism algorithm.
5:  *Identifiable = True*
6:  **for** $\Pi$ in $Gen(Aut(G_P))$ **do**
7:     $G = V\Pi V^\dagger$ ($V^\dagger$ is pseudoinverse)
8:     **if** ($G$ is not signed permutation) **then**
9:         *Identifiable = False* ($P$ is not identifiable)
10:       $break$ the loop
11:     **else**
12:         $continue$
13:     **end if**
14: **end for**

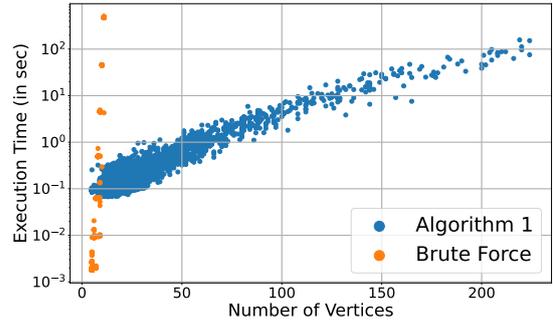

**Fig. 2**: Execution Time of the Algorithms

Figure 2 illustrates the execution time of the algorithm with respect to the number of vertices of the polytope for both Algorithm 1 and the brute-force approach. As expected, the Algorithm 1 scales exponentially with the number of vertices. Remarkably, 96 % of these polytopes are classified as identifiable. Moreover, we include the brute-force algorithm execution time for the polytopes up to 11 vertices in Figure 2. We note that even for polytopes having more than 12 vertices, the identifiability decision with brute-force approach is not feasible since the algorithm complexity scales much faster than Algorithm 1.

In this article, we proposed a computationally tractable polytope identifiability decision algorithm by using graph automorphism and illustrated its efficiency. Although the asymptotic complexity of the graph isomorphism algorithm seems discouraging in the worst-case, our experiments verify that the proposed approach provides a practically feasible solution to the polytope identifiability determination problem.